# Managing Evolving Business Workflows through the Capture of Descriptive Information

Sébastien Gaspard[1, 2, 3], Florida Estrella[1], Richard McClatchey[1] & Régis Dindeleux[2, 3]

[1] CCCS, University of the West of England, Frenchay, Bristol BS16 1QY UK
Richard.McClatchey@cern.ch
[2] LLP/ESIA, Université de Savoie, Annecy, 74016 CEDEX, France
sebastien.gaspard@univ-savoie.fr
[3] Thésame Mécatronique et Management, Annecy, 74000, France
rd@thésame-innovation.com

**Abstract.** Business systems these days need to be agile to address the needs of a changing world. In particular the discipline of Enterprise Application Integration requires business process management to be highly reconfigurable with the ability to support dynamic workflows, inter-application integration and process reconfiguration. Basing EAI systems on model-resident or on a so-called description-driven approach enables aspects of flexibility, distribution, system evolution and integration to be addressed in a domain-independent manner. Such a system called CRISTAL is described in this paper with particular emphasis on its application to EAI problem domains. A practical example of the CRISTAL technology in the domain of manufacturing systems, called Agilium, is described to demonstrate the principles of model-driven system evolution and integration. The approach is compared to other model-driven development approaches such as the Model-Driven Architecture of the OMG and so-called Adaptive Object Models.

## 1. Background and Related Works

As the global marketplace becomes increasingly complex and intricately connected, organizations are constantly pressured to re-organize, re-structure, diversify, consolidate and slim down to provide a winning competitive edge. With the advent of the Internet and e-commerce, the need for coexistence and interoperation with legacy systems and for reduced 'times-to-market', the demand for the timely delivery of flexible software has increased. Couple to this the increasing complexity of systems and the requirement for systems to evolve over potentially extended timescales and the importance of clearly defined, extensible models as the basis of rapid systems design becomes a pre-requisite to successful systems implementation.

One of the main drivers in the object-oriented design of information systems is the need for the reuse of design artefacts or models in handling systems evolution. To be able to cope with system volatility, systems must have the capability of reuse and to adapt as and when necessary to changes in requirements. The philosophy that has been investigated in the research reported in this paper is based on the systematic capture of the description of systems elements covering multiple views of the system to be designed (including processes and workflows) using common techniques. Such a *description-driven approach* [1, 2] involves identifying and abstracting the crucial elements (such as items, processes, lifecycles, goals, agents and outcomes) in the system under design and creating high-level descriptions of these elements which are stored and managed separately from their instances.

Description-driven systems (DDS) make use of so-called *meta-objects* to store domain-specific system descriptions that control and manage the life cycles of meta-object instances or domain objects. The separation of descriptions from their instances allows them to be specified and managed and to evolve independently and asynchronously. This separation is essential in handling the complexity issues facing many web-computing applications and allows the realization of inter-operability, reusability and system evolution as it gives a clear boundary between the application's basic functionalities from its representations and controls. In a description-driven system as we define it, process descriptions are separated from their instances and managed independently to allow the process descriptions to be specified and to evolve asynchronously from particular instantiations of those descriptions. Separating descriptions from their instantiations allows new versions of elements (or element descriptions) to coexist with older versions.

In this paper the development of Enterprise Resource Planning (ERP) in flexible business systems is considered and the need for business process modelling in Enterprise Application Integration (EAI) [3] is established. Workflow systems are considered as vehicles in which dynamic system change in EAI can be catered for as part of handling system evolution through the capture of system description. A description-driven approach is proposed to enable the management of workflow descriptions and an example is given of an application of the CRISTAL description-driven system developed at CERN [4] to handle dynamic system change in workflows. This approach has today two parallel implementations that are called CRISTAL for CMS and Agilium. The two applications are based on the same kernel called CRISTAL KERNEL which provides the DDS functionalities. Both applications inherit these functionalities even if the goals and the specifics of each application is radically different.

## 2. The Need for Integrated Business Process Modelling in EAI

In recent years, enterprises have been moving from a traditional function-led organisation, addressing the needs of a "vertical" market, to a "horizontal" organisation based on business processes. The emergence of new norms such as the ISO 9001 V2000 and the development of inter-enterprise exchanges are further drivers towards process-led reorganisation. However, currently available information systems and production software are still organised following function models. Consequently, they are not well adapted to the exchange of information between enterprises nor to coping with evolving process descriptions. In modern enterprises organised following a horizontal structure, industrial EAI solutions are very dependent on process performance and on the ability of the underlying enterprise management to execute and automate the business processes. Furthermore the requirement for the support of enterprise activities is not only for the execution of internal processes but also for external processes, as in the support of supplier-customer relationships especially in supply chain management.

Enterprise processes have to integrate increasingly more complex business environments including domain-dependent processes for managing both inter-application operation and inter-organisation operation where efficient communications is crucial. Integration sources across enterprises are numerous and multi-technological and can include ERP, human resource management, Customer Relation Management (CRM), administration software, Intranet /Internet, proprietary software and a plethora of office tools.

The first step that an enterprise must make in order to move from a standard vertical organisation to a horizontal organisation is to chart its existing business processes and the interactions between these processes. Following this it must update and manage its internal processes based on existing information systems. For that, the enterprise may be confronted by a collection of different production software systems and their abilities to interact. Most of the software offerings that support ERP deal with the description of enterprises through its organisation by function and examples of these products include systems for purchase service, stock management, production management etc. However individual systems need to synchronise with each other and each normally has their own process management models. Most commercial software do not provide tools to aid in process description and evolution. Even when workflow engines (which can provide synchronisation between systems) are integrated within ERP systems, they are for the most part not synchronised with external applications of the ERP system.

EAI [3] systems concentrate on an architecture centred on software which is dedicated to the interconnection of heterogeneous applications that manage information flows. The heart of EAI software is normally based on the concept of interface references where transformation, routing and domain dependent rules are centralised. Standard EAI architecture is normally built on three layers: processes, routing and transport layers as shown in figure 1.

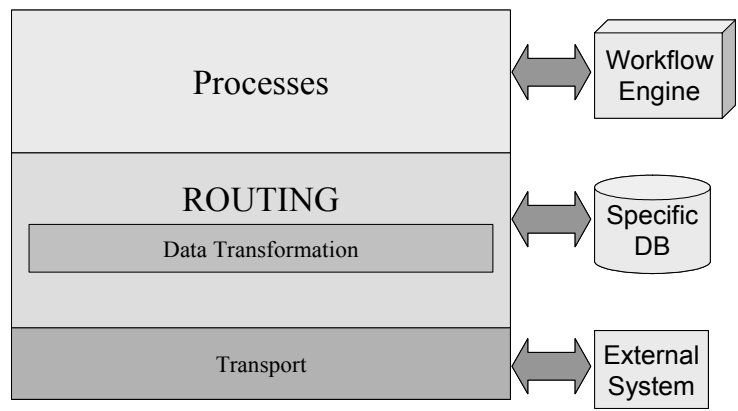

**Figure 1:** The three basic layers of an Enterprise Application Integration (EAI) system.

At the lowest layer are the so-called "connectors" which allow external applications to connect to the EAI platform. It is this level that manages the transport of messages between applications. This Transport layer can be based on basic technologies such as Message Oriented MiddleWare (MOM) [5], and on file reading, email and technologies such as HTTP. The middle layer level of standard EAI software (the Routing layer) manages the transformation of data and its routing between internal systems. More evolved technologies such as XML/XSLT/SOAP, Electronic Data Interchange (EDI), "home-made" connectors and a transition database are used to provide the routing capabilities in this layer. The function of this layer is to apply transformation rules to source application data and to route the new information to the required target application. The third layer of an EAI system is dedicated to system modelling. At this layer, a workflow engine managing domain-dependent specific processes is often employed (when available).

Technically this EAI model suffers from a number of problems:

- The management and modelling of processes needs specific development. Where a workflow engine is used for this purpose, workflows are often based on a monolithic architecture using a matrix definition of workflows that is fixed for the lifecycle of the system.
- Specific technologies are used. In MOM solutions data transformations are not normally based on non-generic tools but on internal developments. Even if XML is used, for most of the case the data dictionary is not defined.
- Guidelines for implementing Connectors do not exist Connectors have to be fully specified, developed and adapted to the connected application. Any change in the EAI software or the connected software requires redevelopment of the Connectors.
- Most of the time, the EAI software has to be placed on a single server, which manages all the processes and has to support three different applications (one for each layer of the EAI model) that have more or less been created to be used together.
- An expensive application server or database management system (DBMS) needs to be already installed and maintained.

As expressed by Joeris in 1999 [6], the support of heterogeneous processes, flexibility, reuse, and distribution are great challenges for the design of the next generation of process modelling languages and their enactment mechanisms. These process modelling technologies are important for business and can make the management of systems more reactive and efficient. However it is not sufficient to concentrate solely on process management and enactment to solve all the problems identified in EAI in the previous section. If workflow systems are not coupled to a comprehensive data management model, optimum functionalities cannot be realised [7]. Most of recent workflow research has been focused on the enactment and modelling of processes based on Petri nets [8], CORBA [9] or UML concepts [10] however when issues of reconfiguration[1] are considered, these research solutions only provides a high level of workflow control that does not completely address the enterprise problems listed above.

The research outlined in this paper proposes an approach that deals with a high level of management of processes with the ability to manage complex data. It is based on distributed technologies and allows the relative autonomy of activity execution, with an enactment model that is relatively similar to that of Joeris [11]. Coupling this technology with some abstraction of process description that can provide generic workflow models [12] is a suitable alternative to standard EAI solutions and more closely addresses the problems listed earlier.

---

[1] Reconfiguration : The ability of a system to dynamically change executing instances of processes in line with a change in its description.

## 3. Handling Evolution via System Description

Approaches for handing system evolution through reuse of design artefacts have led to the study of reusable classes, design patterns, frameworks and model-driven development. Emerging and future information systems however require more powerful data modelling techniques that are sufficiently expressive to capture a broader class of applications. Compelling evidence suggests that the data model must be OO, since that is the model that currently maximises generality. The data model needs to be an open OO model, thereby coping with different domains having different requirements on the data model [13]. We have realised that object *meta-modelling* allows systems to have the ability to model and describe both the static properties of data and their dynamic relationships, and address issues regarding complexity explosion, the need to cope with evolving requirements, and the systematic application of software reuse.

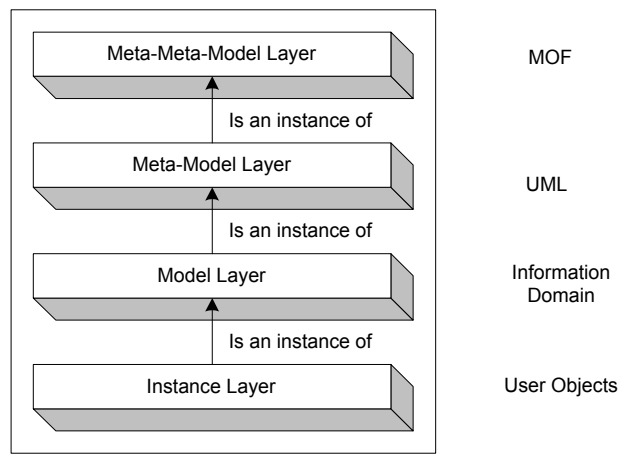

**Figure 2:** The four-layer architecture of the OMG

To be able to describe system and data properties, object meta modelling makes use of meta-data. Meta-data are information defining other data. Figure 2 shows the familiar four-layer model of the Object Management Group, OMG, embodying the principles of meta modelling. Each layer provides a service to the layer above it and serves as a client to the layer below it. The meta-meta-model layer defines the language for specifying meta-models. Typically more compact than the meta-model it describes, a meta-meta-model defines a model at a higher level of abstraction than a meta-model. The meta-model layer defines the language for specifying models, a meta-model being an instance of a meta-meta-model. The model layer defines the language for specifying information domains. In this case, a model is an instance of a meta-model. The bottom layer contains user objects and user data, the instance layer describing a specific information domain. The OMG standards group has a similar architecture based on model abstraction, with the Meta-Object Facility (MOF) model and the UML [14] model defining the language for the meta-meta-model and meta-model layers, respectively.

The judicious use of meta-data can lead to heterogeneous, extensible and open systems. Meta-data make use of a meta-model to describe domains. Our recent research has shown that meta modelling creates a flexible system offering the following - reusability, complexity handling, version handling, system evolution and inter-operability. Promotion of reuse, separation of design and implementation and reification are some further reasons for using meta-models. As such, meta modelling is a powerful and useful technique in designing domains and developing dynamic systems.

A reflective system utilizes an open architecture where implicit system aspects are reified to become explicit first-class meta-objects [15]. The advantage of reifying system descriptions as objects is that operations can be carried out on them, like composing and editing, storing and retrieving, organizing and reading. Since these meta-objects can represent system descriptions, their manipulation can result in change in the overall system behaviour. As such, reified system descriptions are mechanisms that can lead to dynamically evolvable and reusable systems. Meta-objects, as used in the current work, are the self-representations of the system describing how its internal elements can be accessed and manipulated. These self-representations are causally connected to the internal structures they represent i.e. changes to these self-representations immediately affect the underlying system. The ability to dynamically augment, extend and redefine system specifications can result in a considerable improvement in flexibility. This leads to dynamically modifiable systems, which can adapt and cope with evolving requirements.

There are a number of OO design techniques that encourage the design and development of reusable objects. In particular design patterns are useful for creating reusable OO designs [16]. Design patterns for structural, behavioural and architectural modelling have been well documented elsewhere and have provided software engineers with rules and guidelines that they can (re-)use in software development. Reflective architectures that can dynamically adapt to new user requirements by storing descriptive information which can be interpreted at runtime have lead to so-called Adaptive Object Models [17]. These are models that provide meta-information about domains that can be changed on the fly. Such an approach, proposed by Yoder, is very similar to the approach adopted in this paper

A Description-Driven System (DDS) architecture [1, 2], as advocated in this paper, is an example of a reflective meta-layer (i.e. meta-level and multi-layered) architecture. It makes use of meta-objects to store domain-specific system descriptions, which control and manage the life cycles of meta-object instances, i.e. domain objects. The separation of descriptions from their instances allows them to be specified and managed and to evolve independently and asynchronously. This separation is essential in handling the complexity issues facing many web-computing applications and allows the realization of inter-operability, reusability and system evolution as it gives a clear boundary between the application's basic functionalities from its representations and controls. As objects, reified system descriptions of DDSs can be organized into libraries or frameworks dedicated to the modelling of languages in general, and to customizing its use for specific domains in particular. As a practical example of our approach the next section describes the DDS architecture developed in the context of research carried out in the CRISTAL project at CERN and the Agilium project at Thesame.

## 4. CRISTAL - a Description-Driven System (DDS)

The Compact Muon Solenoid (CMS) is a general-purpose experiment at CERN that will be constructed from around a million parts and will be produced and assembled in the next decade by specialized centres distributed worldwide. As such, the construction process is very data-intensive, highly distributed and ultimately requires a computer-based system to manage the production and assembly of detector components. In constructing detectors like CMS, scientists require data management systems that are able to cope with complexity, with system evolution over time (primarily as a consequence of changing user requirements and extended development timescales) and with system scalability, distribution and interoperation.

No commercial products provide the capabilities required by CMS. Consequently, a research project, entitled CRISTAL (Cooperating Repositories and an Information System for Tracking Assembly Lifecycles [4]) has been initiated to facilitate the management of the engineering data collected at each stage of production of CMS. CRISTAL is a <u>distributed product data and workflow management system</u>, which makes use of an OO database for its repository, a multi-layered architecture for its component abstraction and dynamic object modelling for the design of the objects and components of the system. CRISTAL is based on a DDS architecture using meta-objects. The DDS approach has been followed to handle the complexity of such a data-intensive system and to provide the flexibility to adapt to the changing scenarios found at CERN that are typical of any research production system. In addition CRISTAL offers domain-independence in that the model is generic in concept. Lack of space prohibits further discussion of CRISTAL; detail can be found in [1, 2 & 4].

The design of the CRISTAL prototype was dictated by the requirements for adaptability over extended timescales, for system evolution, for interoperability, for complexity handling and for reusability. In adopting a description-driven design approach to address these requirements, the separation of object instances from object description instances was needed. This abstraction resulted in the delivery of a three layer description-driven architecture. The model abstraction (of instance layer, model layer, meta-model layer) has been adapted from the OMG MOF specification [18], and the need to provide descriptive information, i.e. meta-data, has been identified to address the issues of adaptability, complexity handling and evolvability.

Figure 3 illustrates the CRISTAL architecture. The CRISTAL model layer is comprised of class specifications for CRISTAL type descriptions (e.g. PartDescription) and class specifications for CRISTAL classes (e.g. Part). The instance layer is comprised of object instances of these classes (e.g. PartType#1 for PartDescription and Part#1212 for Part). The model and instance layer abstraction is based on model abstraction and on the *Is an instance of* relationship. The abstraction based on meta-data abstraction and the *Is described by* relationship leads to two levels - the meta-level and the base level. The meta-level is comprised of meta-objects and the meta-level model that defines them (e.g. PartDescription is the meta-level model of PartType#1 meta-object). The base level is comprised of base objects and the base level model that defines them (Part is the base-level model of Part#1212 object).

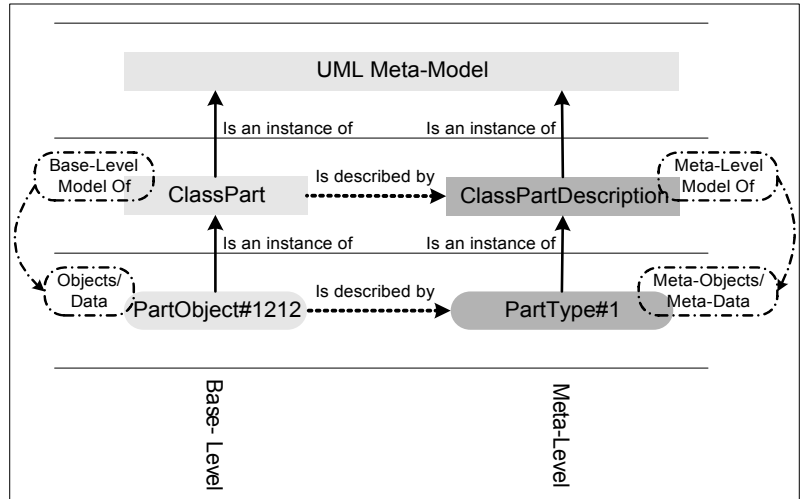

**Figure 3:** The CRISTAL description-driven architecture.

The approach of reifying a set of simple design patterns as the basis of the description-driven architecture for CRISTAL has provided the capability of catering for the evolution of a rapidly changing research data model. In the two years of operation of CRISTAL it has gathered over 25 Gbytes of data and been able to cope with more than 30 evolutions of the underlying data schema without code or schema recompilations.

## 5. Agilium - a Description Driven Workflow System

### 5.1 Agilium Functionality

In order to address the deficiencies in current EAI systems a research system entitled Agilium, based on CRISTAL technologies, has been developed by a collaboration of three research partners, CERN (the European Organisation for Nuclear Research, in Geneva Switzerland), UWE (the University of the West of England, Bristol UK) and Thésame (an innovation network for technology companies in mechatronic, computer-integrated manufacturing and management, based in Annecy France). The model and technologies used in Agilium make EAI tools accessible to middle-sized enterprises and to software houses and integrators that target the EAI market. The CRISTAL architecture is providing a coherent way to replace the three application layers of EAI (as shown in figure 1) by a single generic layer, based on Items and using common tools, processes, routing and transport. In order to provide an effective EAI architecture, Agilium combines Items described using the DDS philosophy. This approach provides a development free way to provide EAI functionality whilst managing behaviour through workflows. Items can be connectors or conceptual domain-specific objects such as *order forms, supplies, commands etc*.

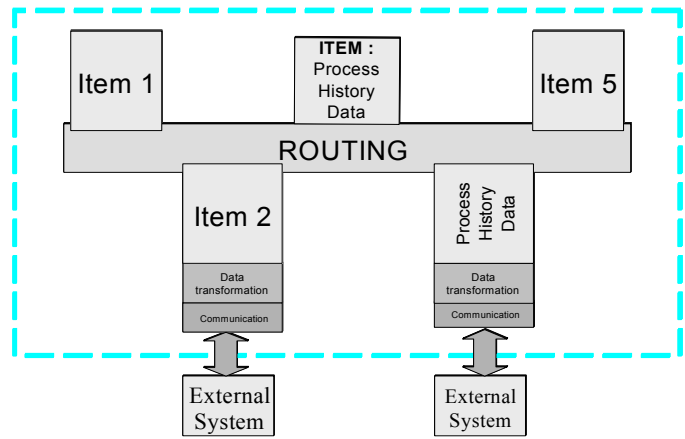

**Figure 4:** The CRISTAL approach to integrated EAI employed in Agilium.

In the Agilium system, a connector is managed as one single Item described with a specific graphically represented behaviour. A connector can transform data (using scripting or XML technologies) and is coupled to a communication method that can be any appropriate enabling technology. In this way, it is easy to connect applications that have any arbitrary communication interface. It is sufficient simply to describe a communication mode (CORBA, HTTP, SOAP, Web Service, text file, email...), a data format (will be converted in XML) and a behaviour (with the workflow graphical interface). Basing on the concept of the DDS, connectors are easily maintainable and modifiable and make the Agilium system easy to integrate and adapt to evolving environments prevalent in enterprise information systems. By combining the Items describing domain specific functionality and those that can connect external applications, the EAI architecture is complete and presents all the functionalities of the external architectures, and more.

Using the facilities for description and dynamic modification in CRISTAL, Agilium is able to provide a modifiable and reconfigurable workflow. The workflow description and enactment elements of CRISTAL are correlated and each instance stores any modifications that have been carried on it. With an efficient model of verification at both levels (i.e. description and enactment) it is possible to validate if the migration from one description to another one within an instance is possible and to detect any modifications and changes and therefore to apply the migration. Ongoing research is being conducted to mathematically model the workflow concepts that could be directly applied to the CRISTAL technologies so as to complete the modification ability.

**5.2 Advantages and limitations of Agilium**

Innovating technologies used in the kernel of CRISTAL provide Agilium significant advantages when compared to standard EAI solutions:

- Flexibility. Architecture independence allows Agilium to adapt to new domains and/or new enterprises without any specific development. This is an essential factor in helping to reduce maintenance costs and to minimise conversation costs, thus providing high levels of flexibility.
- Platform independence. Use of JAVA/CORBA/XML/LDAP technologies allows CRISTAL to work on any preinstalled operating system (Linux, Windows, UNIX, Mac OS) and on any machine (Mac, PC, Sun, IBM...).
- Database independence. XML storage with LDAP referencing makes CRISTAL autonomous and independent of any underlying database and of any application server.
- Simplified integration of applications. XML is becoming the standard for interfacing applications within an enterprise. It presents a solution that supports multiple heterogeneous environments. With the limitation of the development of a translation/transport layer, connectors are based on a generic model.
- Fully distributed. This functionality provides web usability through the Internet or an Intranet. It also makes data accessible from multiple databases via a unique interface.
- CRISTAL's powerful workflow management facilities provide the ability to model and execute processes of any type that can also evolve by dynamic modification or appliance of any new description.

But there are some limitations:
- Because it is based on graphical descriptions, it is not always simple to determine the way to code (describing a workflow) CRISTAL actions. Using a high level of abstraction can render simple things difficult to represent in code.
- Providing complete flexibility to users to define elements in the system can compromise the integrity of enactment and the implementation is not yet sufficiently advanced to provide a fully secured system without requiring human intervention.
- As the connectivity technologies such as BPML and BPEL4W [19] become more complex, complete, normative and numerous, the Agilium tool has to provide and maintain many connectors that can not only be defined by the user.

### 5.3 Future work

Ongoing research is being conducted into the mathematical approach of process modelling in CRISTAL which may ultimately include a decision-making model based on agent technology. It is planned that these agents would verify changes to the model and dynamically modify the instances of the workflow processes basing its calculation and decision on a base of user predefined constraints that must be respected. Another aspect that may be explored is to use an Architecture Definition Language to model Items and their interactions. This would provide an efficient and secure way to create descriptions for new domains to model. Another area that could be explored is the connector aspects for the Agilium EAI. A mathematical approach of connector specifications could be envisaged and defined which would enable connector development to be automated in CRISTAL.

## 6. Conclusions

The combination of a multi-layered meta-modelling architecture and a reflective meta-level architecture resulted in what has been referred to in this paper as a DDS architecture. A DDS architecture, is an example of a reflective meta-layer architecture. The CRISTAL DDS architecture was shown to have two abstractions. The vertical abstraction is based on the *is an instance of* relationship from the OMG meta-modelling standard, and has three layers - instance layer, model layer and meta-model layer. This paper has proposed an orthogonal horizontal abstraction mechanism that complements this OMG approach. The horizontal abstraction is based on the meta-level architecture approach, encompasses the *is described by relationship* and has two layers - meta-level and base level. This description-driven philosophy facilitated the design and implementation of the CRISTAL project with mechanisms for handling and managing reuse in its evolving system requirements and served as the basis of the Agilium description-driven workflow system.

The model-driven philosophy expounded in this paper is similar to that expounded in the Model Driven Architecture (MDA [18]) of the OMG. The OMG's goal is to provide reusable, easily integrated, easy to use, scalable and extensible components built around the MDA. While DDS architectures establish those patterns, which are required for exploiting data appearing at different modelling abstraction layers, the MDA approaches integration and interoperability problems by standardizing interoperability specification at each layer (i.e. standards like XML, CORBA, .NET, J2EE). The MDA integration approach is similar to the Reference Model for Open Distributed Processing (RM-ODP) [20] strategy of interoperating heterogeneous distributed processes using a standard interaction model. In addition, the Common Warehouse Metamodel (CWM) specification [21] has been recently adopted by the OMG. The CWM enables companies to manage their enterprise data better, and makes use of UML, XML and the MOF. The specification provides a common meta-model for warehousing and acts as a standard translation for structured and unstructured data in enterprise repositories, irrespective of proprietary database platforms.

Likewise, the contributions of this work complement the ongoing research on Adaptive Object Model (AOM) espoused in [17] and [22], where a system with an AOM (also called a Dynamic Object Model) is stated to have an explicit object model that is stored in the database, and interpreted at runtime. Objects are generated dynamically from the AOM schema meta-data that represent data descriptions. The AOM approach also uses reflection in reifying implicit data aspects (e.g. database schema, data structures, maps of layouts of data objects, references to methods or code). The description-driven philosophy has demonstrably facilitated the design and implementation of the CRISTAL and Agilium projects with mechanisms for handling and managing reuse in its evolving system requirements.

## Acknowledgments

The authors take this opportunity to acknowledge the support of their home institutes and numerous colleagues responsible for the CRISTAL & Agilium software.